\documentclass[sigconf, twocolumn]{acmart}

\usepackage{xcolor}

\usepackage{listings}
\usepackage{xcolor}
\usepackage{setspace}
\usepackage{geometry}

\definecolor{codegreen}{rgb}{0,0.6,0}
\definecolor{codegray}{rgb}{0.5,0.5,0.5}
\definecolor{codepurple}{rgb}{0.58,0,0.82}
\definecolor{backcolour}{rgb}{0.95,0.95,0.92}

\lstdefinestyle{mystyle}{
    commentstyle=\color{codegreen},
    keywordstyle=\color{magenta},
    numberstyle=\tiny\color{codegray},
    stringstyle=\color{codepurple},
    basicstyle=\ttfamily\footnotesize,
    breakatwhitespace=false,         
    breaklines=true,                 
    captionpos=t,                    
    keepspaces=true,                 
    numbers=left,                    
    numbersep=5pt,                  
    showspaces=false,                
    showstringspaces=false,
    showtabs=false,                  
    tabsize=2
}

\AtBeginDocument{%
  \providecommand\BibTeX{{%
    \normalfont B\kern-0.5em{\scshape i\kern-0.25em b}\kern-0.8em\TeX}}}

\definecolor{gray50}{gray}{.5}
\definecolor{gray40}{gray}{.6}
\definecolor{gray30}{gray}{.7}
\definecolor{gray20}{gray}{.8}
\definecolor{gray10}{gray}{.9}
\definecolor{gray05}{gray}{.95}

\definecolor{celadon}{rgb}{0.67, 0.88, 0.69}
\definecolor{lightcoral}{rgb}{0.94, 0.5, 0.5}

\definecolor{black}{rgb}{0.2, 0.2, 0.2}
\definecolor{darkgreen}{rgb}{0, 0.5, 0}

\definecolor{lightgrey}{rgb}{0.9, 0.9, 0.9}

\newcommand{\PreserveBackslash}[1]{\let\temp=\\#1\let\\=\temp}
\makeatletter
\newcommand\footnoteref[1]{\protected@xdef\@thefnmark{\ref{#1}}\@footnotemark}

\makeatother

\usepackage{array}
\newcolumntype{P}[1]{>{\centering\arraybackslash}p{#1}}

\newcommand{\Comment}[1]{} 

\definecolor{dkgreen}{rgb}{0,0.6,0}
\definecolor{gray}{rgb}{0.5,0.5,0.5}
\definecolor{mauve}{rgb}{0.58,0,0.82}

\definecolor{mygray}{gray}{1}

\setcopyright{acmlicensed}
\copyrightyear{2024}
\acmYear{2024}
\acmDOI{XXXXXXX.XXXXXXX}

\acmConference[SBES'24]{Brazilian Symposium on Software Engineering}{September 30 -- October 04, 2024}{Curitiba, PR}
\acmBooktitle{Brazilian Symposium on Software Engineering (SBES ’24),
 September 30-October 4, 2024, Curitiba, Brazil} 

\acmISBN{978-1-4503-XXXX-X/18/06}

\begin{document}

\definecolor{cleidson}{cmyk}{0,100,100,0} 
\newcommand{\cleidson}[1]{\textcolor{cleidson}{#1}}

\title{Assisting Novice Developers Learning in Flutter Through Cognitive-Driven Development}

\author{Ronivaldo Ferreira}
\orcid{https://orcid.org/0000-0002-5812-8707}
\affiliation{%
  \institution{Federal University of Pará}
  \streetaddress{R. Augusto Corrêa, 01 - Guamá, Belém - PA, 66075-110}
  \city{Belém}
  \state{Pará}
  \country{Brazil}
}
\email{ronivaldo.junior@icen.ufpa.br}

\author{Victor Hugo Santiago C. Pinto}
\orcid{https://orcid.org/0000-0001-8562-6384}
\affiliation{%
  \institution{Federal University of Pará}
  \streetaddress{R. Augusto Corrêa, 01 - Guamá, Belém - PA, 66075-110}
  \city{Belém}
  \state{Pará}
  \country{Brazil}
}
\email{victor.santiago@ufpa.br}

\author{Cleidson R. B. de Souza}
\orcid{0000-0003-3240-3122}
\affiliation{%
  \institution{Federal University of Pará}
  \city{Belém, Pará}
  \country{Brazil}}
\email{cleidson.desouza@acm.org}

\author{Gustavo Pinto}
\orcid{https://orcid.org/0000-0001-7598-2799}
\affiliation{%
  \institution{Federal University of Pará \& Zup Innovation}
  \streetaddress{R. Augusto Corrêa, 01 - Guamá, Belém - PA, 66075-110}
  \city{Belém}
  \state{Pará}
  \country{Brazil}
}
\email{gpinto@ufpa.br}

\renewcommand{\shortauthors}{Ferreira, R.; Pinto, V. H. S.; C., R. B. de Souza and Pinto, G.}

\begin{abstract}

Cognitive-Driven Development (CDD) is a coding design technique that helps developers focus on designing code within cognitive limits. The imposed limit tends to enhance code readability and maintainability. While early works on CDD focused mostly on Java, its applicability extends beyond specific programming languages. In this study, we explored the use of CDD in two new dimensions: focusing on Flutter programming and targeting novice developers unfamiliar with both Flutter and CDD. Our goal was to understand to what extent CDD helps novice developers learn a new programming technology.
We conducted an in-person Flutter training camp with 24 participants. After receiving CDD training, six remaining students were tasked with developing a software management application guided by CDD practices. Our findings indicate that CDD helped participants keep code complexity low, measured using Intrinsic Complexity Points (ICP), a CDD metric. Notably, stricter ICP limits led to a 20\% reduction in code size, improving code quality and readability. This report could be valuable for professors and instructors seeking effective methodologies for teaching design practices that reduce code and cognitive complexity.
\end{abstract}

\keywords{Cognitive-Driven Development, Software Design}


\received{20 February 2007}
\received[revised]{12 March 2009}
\received[accepted]{5 June 2009}

\maketitle

\section{Introduction}
The need for better software design techniques is crucial to support the smooth evolution of a software project~\cite{parnas1994software}. By providing principled abstractions and controlling code complexity, these design techniques serve as valuable guides to developers during their coding process. Consequently, previous studies have introduced several techniques to ease the maintainability of a software codebase~\cite{martin2000design, hutton2009clean, evans2004domain, fraser2003test}.

Despite the importance of these design techniques in reducing software complexity, it is still common to find software products suffering from design decay, a scenario in which ``developers progressively introduce code with poor design structures into a system'''~\cite{barbosa2020revealing}, making code coprehension, maintenance, and evolution more difficult. Understanding design techniques is particularly relevant for novice developers, as their lack of experience might lead them to learn from both good and bad experiences directly from the codebase. Therefore, to help students understand and navigate software complexities, professors and instructors should incorporate design practices into the software engineering curriculum.

One reason these design approaches have not become dominant tools in the programmer community is their subjectivity. Take, for instance, the Single-Responsibility Principle, one of the SOLID practices. This principle suggests that a class should have one and only one reason to change. Consider an application that calculates the sum of areas for a collection of shapes (circles and squares). The \texttt{AreaCalculator} class handles this logic, but it also includes the responsibility of outputting the data. Thus, the mixing of responsibilities can be subjective—some developers might prefer separating these concerns.

Cognitive-Driven Development (CDD) is a design approach that removes subjectiveness by introducing Intrinsic Complexity Points (ICP), which are code elements that could affect developers’ understanding according to their usage frequency. By defining and following a set of ICPs, developers can control the complexity in a given code unit~\cite{pinto2023cognitive}. Early works on CDD observed that the technique can be useful in designing modular and maintainable code~\cite{barbosa2022extent,pinto2021cognitive}. 
However, these research efforts have focused on skilled Java programmers. Consequently, little is known about the effectiveness of CDD for either 1) less experienced developers or 2) distinct programming languages.

This study aims to address these gaps by observing how novice developers unfamiliar with CDD utilize it to build a management system application using Flutter, a popular Dart web/mobile framework. To achieve this goal, we conducted an in-person workshop, attended by 24 students who learned about CDD code principles. After the training, teams of six students were formed to build a mobile application.


To guide our study, we designed the following research questions:

\begin{itemize}
    \item[RQ1] How does CDD aid novice developers to understand and manage software complexity?
    \item[RQ2] How do novice developers perceive the effectiveness of CDD?
    \item[RQ3] What were the challenges for new developers in understanding and utilizing CDD effectively?
\end{itemize}

These questions aim to evaluate the use of CDD as a tool for helping novice developers learn and understand software design. Among the findings, we observed that the use of CDD helped novice developers focus on the most challenging areas of learning, while allowing them to assess their progress across different cycles of the process. We also noticed that participants positively accepted the use of CDD, resulting in better engagement and dynamics. Finally, it was possible to measure the practical impact of CDD; in a CDD-guided refactoring activity, the team achieved a 20\% decrease in the total number of lines of code.
Our findings reinforce the usefulness of CDD, not only as a design technique that helps developers design less complex code units but also as a tool for learning and managing code complexity. Our artifacts are publicly accessible at~\cite{artifacts}.



\section{A brief introduction to CDD}

CDD (Cognitive-driven Development) is a software development approach proposed by Souza and Pinto~\cite{de2020toward} that aims to reduce effort in the software development process by limiting the number of programming constructs that developers could incorporate into their code units. This enables developers to map and control these programming elements, identifying cognitive overloads.

CDD is primarily based on the research of George A. Miller~\cite{miller1956magical} and John Sweller~\cite{sweller1988cognitive, sweller2010element}. Miller discovered that individuals tend to retain about seven units of information in their short-term memory, with a variation of approximately two units. Sweller’s Cognitive Load Theory (CLT) focuses on the amount of information we can process simultaneously, highlighting the challenge of dealing with cognitive overload in situations that require short-term memory storage. He emphasizes that some materials are inherently complex, making them difficult to comprehend.

Broadly, CDD proposes setting a complexity limit, establishing complexity points in code to identify when refactoring is needed due to cognitive overload. This involves designating metrics of cognitive complexity for specific code components. Intrinsic Complexity Points (ICPs) quantify the inherent complexity of elements in the source code. These indicators represent the intrinsic complexity of the analyzed code, helping to differentiate the impact of each code structure among developers. Consequently, throughout software evolution, code units tend to become more concise and readable, reducing maintenance costs~\cite{pinto2023cognitive}.

CDD was the focus of several research works. For instance, Pinto et al.~\cite{pinto2021cognitive} analyzed the application of CDD in the refactoring process using object-oriented metrics. Pinto and Souza~\cite{iceis22} investigated the impacts of implementing CDD in the initial coding stages compared to traditional practices. Barbosa et al.~\cite{barbosa2022extent} evaluated the use of CDD to enhance code readability. Pinto and Souza~\cite{pinto2023cognitive} shared insights on building an industry tool from scratch using CDD. Pereira et al.~\cite{pereira2021cognitive} presented Cognitive Load Analyzer, a noteworthy tool supporting the adoption of CDD.

These previous works on CDD focused on a very specific population of skilled Java developers~\cite{de2020toward, pereira2021cognitive, pinto2021cognitive, iceis22, pinto2023cognitive}. It is unclear though how these benefits translate in a scenario where developers are still learning programming while managing software complexity --- which is the goal of this work.


\section{PROPOSAL OF ICPs FOR FLUTTER}

This section presents a proposal for ICPs. We started from the recommendations of previous studies~\cite{de2020toward, pinto2021cognitive, pereira2021cognitive, iceis22, barbosa2022extent, pinto2023cognitive} to formulate an initial set of ICPs. These suggestions facilitated the construction of the team's ICPs (see section 4.4).

{\subsection{Why Other ICPs?}

Dart is a programming language designed by Google in 2011. Flutter is a Dart's framework that focus on mobile and web development. Given the design of Dart/Flutter, it becomes important for this study to derive its own set of ICPs. This happens because Dart/Flutter have different programming constructs than Java-like languages. For instance, since Dart/Flutter were designed specifically for building modern, high-performance user interfaces, the declarative approach in Flutter enables developers to construct complex UIs with minimal code, promoting reusability and maintainability. Additionally, Dart's language features, such as optional typing and async-await, streamline development processes and improve code readability, making it particularly suited for UI-centric applications, and is less verbose than Java's concurrency model.
As a consequence, ICPs adopted in languages like Java cannot be easily reused, highlighting the need of defining new ICPs for Dart/Flutter.


\subsection{Suggested ICPs}

Initially, we were inspired by the ICPs used in the work of Souza and Pinto~\cite{de2020toward}. Then we added elements of Dart/Flutter. These elements were chosen based on the instructors experience and the observation of the students learning curve during the training camp. Table~\ref{tab:suggested_icps} list the ICPs designed.

\begin{table}[h!]
\centering
\small
\caption{Collection of suggested ICPs.}
\label{tab:suggested_icps}
\begin{tabular}{@{}>{\centering\arraybackslash}p{0.32\linewidth}p{0.48\linewidth}>{\centering\arraybackslash}p{0.10\linewidth}c@{}}
\toprule
\textbf{Category} & \multicolumn{1}{c}{\textbf{Description}} & \textbf{Weight}
\\ \midrule
Branches and Loops & Control structures such as \texttt{if}, \texttt{else}, \texttt{for}, \texttt{while}, \texttt{case}, and ternary operators & 1 \\
Coupling & Usage of functions as arguments and dependencies on external components or services & 1 \\
Nullable & Handling of nullable widgets and variables & 1 \\
Asynchronous Function & Implementation of asynchronous operations using \texttt{Future} and \texttt{Stream} & 2 \\
Asynchronous Widget & Utilization of asynchronous widgets like \texttt{FutureBuilder} and \texttt{StreamBuilder} & 2 \\
State Management & Management of state using libraries like \texttt{Provider} and \texttt{Flutter\_bloc} & 2 \\
Animated Widget & Creation of animations, both implicit/explicit and low-level animations & 2 \\
\bottomrule
\end{tabular}
\end{table}

As one can see, we have ICPs for both Dart and Flutter. We will describe each one next.

\vspace{0.2cm}
\noindent
\textbf{Branches and Loops (Dart)}
The use of branches and loops is common programming feature. 

\vspace{0.2cm}
\noindent
\textbf{Coupling (Dart)}
Composition involves combining simpler components to create complex ones, promoting reuse. Thus, composing more complex and customized elements through coupling plays a role in the modularity of the code.

\vspace{0.2cm}
\noindent
\textbf{Nullable (Dart)}
Managing null values can increase code complexity, particularly in situations where nullability is not explicitly. Determining when and how to incorporate nullability can present an additional challenge for novice developers.

\vspace{0.2cm}
\noindent
\textbf{Asynchronous Function (Dart)}
Asynchronous functions are essential for maintaining the responsiveness of the user interface during long-running operations, such as network or background processing. They allow the application to continue responding to user events, enhancing user experience. 

\vspace{0.2cm}
\noindent
\textbf{Asynchronous Widget (Flutter)}
When the application needs to perform time-consuming operations, asynchronous widgets ensure that the user interface remains responsive. However, the concept of Future and asynchronous operations can be hard for novice developers to understand and eventually manage errors that may occur during Future execution.

\vspace{0.2cm}
\noindent
\textbf{State Management (Flutter)}
State management in Flutter is essential but complex for beginners. The variety of concepts such as streams and blocs, along with the different available libraries, creates uncertainties in choosing the ideal approach. 

\vspace{0.2cm}
\noindent
\textbf{Animated Widget (Flutter)}
Well-planned animations enhance the user interface, adding sophistication to the application. Flutter offers extensive support with various animation approaches, including predefined motion effects that can be customized as needed, or even creating animations by drawing widgets frame by frame.

\subsection{Weights and Limits of ICPs}

To establish the limit of ICPs, Souza and Pinto~\cite{de2020toward} recommend restricting the number of ICPs per code unit to between 3 and 7 for teams with different levels of experience, and between 10 and 12 for high performing teams. 

In addition to the number of ICPs in a code unit, we also considered the weight of each ICP. For instance, we assumed that ICPs related to mobile/web development are more challenging for novice developers to understand compared to traditional ICPs like branches and loops. 
Taking the ``Asynchronous Widget'' category as an example, novice developers may face various concerns, for example, understanding asynchronous operations through the Future object, managing widget states, and error handling. Figure~\ref{fig:example-1} illustrates an example of code adapted from the Flutter documentation, which uses the \texttt{FutureBuilder} to manage an asynchronous operation.

\begin{figure}
\begin{lstlisting}[language=Python,framerule=1pt, linewidth=7.8cm, basicstyle=\scriptsize]
FutureBuilder(
  future: someAsyncFunction,
  builder: (context, snapshot) {
    if (snapshot.connectionState == ConnectionState.waiting) 
      return CircularProgressIndicator();
    if (snapshot.hasData) 
      return _buildBodyComponet();
    if (snapshot.hasError) 
      return _buildErrorComponent();
    return ...;
  }
)

\end{lstlisting}
\caption{Code example for the ICP in the “Asynchronous Widget” category.}
\label{fig:example-1}
\end{figure}

In the ``builder'' parameter, the user interface is dynamically controlled based on the connection state of the Future. If the state is \texttt{ConnectionState.waiting}, a progress indicator, such as a \texttt{CircularProgressIndicator}, is displayed, providing visual feedback to the user while the asynchronous operation is running. Utilizing the \texttt{AsyncSnapshot} class, received data is manipulated to construct the user interface responsively. When data is available (\texttt{snapshot.hasData}), a set of corresponding custom widgets is rendered. Conversely, error detection (\texttt{snapshot.hasError}) facilitates proper management of error situations, ensuring the reliability of the application.

A weight of 2 was assigned to the ICP of the ``Asynchronous Widget'' category, due to developers' concerns regarding the understanding of Future, with could impact on the maintenance and usability of the system. 
Generally speaking, most Dart-related ICPs were given a weight of ``1,'' while most Flutter-related ICPs were assigned a weight of ``2.'' Therefore, the limit of ICPs was determined not just by the count of their occurrences in code units but also by their weights, taking into account the complexities faced by developers.

\section{In-Person Workshop}

For this work, we conducted an in-person workshop focused on learning mobile development using the Flutter framework. Held at the Faculty of Computing (FACOMP) of the Federal University of Pará (UFPA), the workshop was divided into three stages: 1) participant selection, 2) developer traning, and 3) practical implementation.
The workshop had 107 hours of workload for those participants who completed all activities. It happened from July to September 2023. The workshop served as a starting point to apply the CDD approach and assess its impacts on novice developers.

\begin{figure*}
    \centering
    \includegraphics[width=0.75\linewidth]{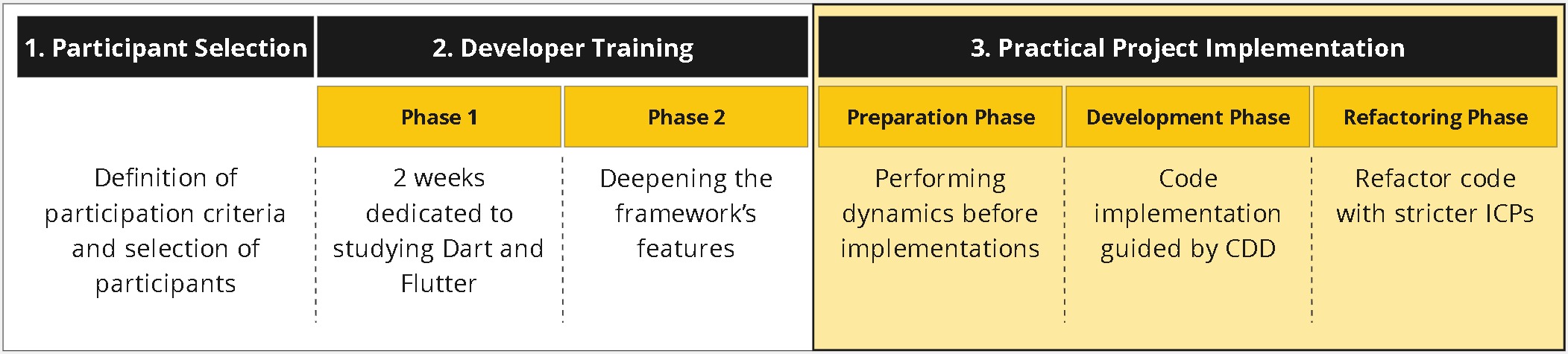}
    \caption{Workshop overview}
    \label{fig:storytelling}
\end{figure*}

Figure~\ref{fig:storytelling} provides an overview of the three stages of the workshop. In the first stage, participant selection took place, establishing criteria for their inclusion. The second stage focused on developer training, divided into two other phases: 1) for studying Dart and Flutter and 2) for deepening the knowledge of the specific technologies of the framework. The third stage was focused on the implementation of the practical project, guided by CDD. In this phase, participants were guided through three different phases, covering initial preparation regarding the necessary methods and tools, the development of application functionalities, and the introduction to refactoring activities.

\subsection{Participant Selection}
For the participant selection stage, students from Computer Science, Computer Engineering, Information Systems, or related fields were sought, both at UFPA and other higher education institutions in the region. Students had to have basic understanding in algorithms and programming.
Participation in the program required a commitment of 2 to 8 weeks, with 15 hours per week dedicated to each stage. Moreover, it was essential to be available for all activities in the training and practical project implementation stages.
Participant registration was made available through online forms. 
To ensure effective participation, candidates needed to meet all the previously detailed requirements, which were extensively communicated during the promotion and registration process.

\subsection{Developer Training}
The developers' training stage aimed to provide both theoretical and practical foundations for building cross-platform applications using the Flutter framework. It was divided into two distinct phases. In Phase 1, one week was dedicated to learning Dart, followed by another week focused on learning Flutter. This phase had 20 participants in theoretical and practical exercises over five classes, totalling 30 hours.
In the second phase, participants were divided into two groups: one dedicated to web technology and the other to mobile resources. The objective of this phase was to deepen participants' knowledge of software development techniques specific to their chosen platform. Phase 2 had 14 participants (six drop out) 
who had 21 hours of workload, including in-class and asynchronous tasks, culminating in individual evaluations.
All material used during this stage was based on Dart documentation~\cite{dart_docs}, Flutter documentation~\cite{flutter_docs}, and Marco L. Napoli's book "Beginning Flutter: A Hands-On Guide To App Development"~\cite{napoli2019beginning}. To deepen the participants' understanding of the CDD approach, we gave a lecture about CDD, proving links for blogs, interviews and research paper as support material.

\subsection{Practical Implementation}
This stage was used to observe the 
results generated from using CDD 
as a guide to control the complexity of the source code during its evolution.

\subsubsection{Application}
The CITIAmazon laboratory aims to promote research and development in the field of market-oriented technologies, establishing a connection between academic research and societal demands. Given this scenario, it became necessary to create management system for this laboratory.
The application not only aims to efficiently manage projects associated with the laboratory but also seeks to track students, scholarship holders, and other associates through actions aimed at the development of their professional careers, as well as promoting community engagement by connecting them with local companies.

\subsubsection{Phases}
This stage was sub-divided into three ones: Preparation Phase, Development Phase, and Refactoring Phase. During Preparation Phase, developers dedicated themselves to methodological preparations for subsequent activities. Development Phase corresponded to the period when coding tasks were performed. Finally, in the Refactoring Phase, developers focused their efforts on enhancing the previously developed code. Table~\ref{tab:schedule} presents the detailed schedule of this stage.

\begin{table}[ht]
\centering
\small
\caption{Project Implementation schedule.}
\label{tab:schedule}
\begin{tabular}{@{}p{0.2\linewidth}p{0.2\linewidth}p{0.52\linewidth}@{}}
\toprule
\textbf{Date}       & \textbf{Phases}             & \textbf{Activities}                                   \\ \midrule
Aug 28 to Sep 1 & Preparation & Interview with client, requirements definition, and backlog construction \\
Sep 4 to Sep 30 & Development & Software Development, Scrum Ceremonies and CDD Tasks    \\
Oct 3 to Oct 10 & Refactoring & Refactoring and CDD Tasks              \\ \bottomrule
\end{tabular}
\end{table}

The initial phase of the stage began with an exploration of the application requirements. Subsequently, software implementation unfolded across four sprints, encompassing development tasks, Scrum ceremonies, and CDD activities. Later, the Refactoring phase was dedicated to enhancing code quality through the application of CDD principles for refactoring.
This schedule highlights the application of an agile and iterative approach. In addition, CDD was used throughout the project. 

\subsubsection{Team Members}
The practical project implementation stage began with a development team of six participants --- the other participants decided not to participate in this phase. Each team member is identified as TM\emph{X}.
Table~\ref{tab:demographics} presents an overview of the team's demographic characteristics. As one can see, the majority of team members are in the early stages of their careers, with experience ranging from 7 months to 3 years, averaging approximately 1 year and a half.

\begin{table*}[ht]
\centering
\small
\caption{Team member demographics. CS, CE, and IS stand for Computer Science, Computer Engineering, and Information Systems, respectively.}
\label{tab:demographics}
\begin{tabular}{@{}p{0.18\linewidth}p{0.11\linewidth}p{0.11\linewidth}p{0.11\linewidth}p{0.11\linewidth}p{0.11\linewidth}p{0.11\linewidth}@{}}
\toprule
\textbf{Date} & \textbf{TM1} & \textbf{TM2} & \textbf{TM3} & \textbf{TM4} & \textbf{TM5} & \textbf{TM6} \\ \midrule
Age & 17 & 23 & 27 & 23 & 18 & 28 \\
Gender & M & M & M & M & M & F \\
Course & CS & CS & CE & IS & IS & CE \\
Programming experience & 1 year & 1 year & 2 years & 7 months & 2 years & 3 years \\
Main prog. language & Python & Python & Javascript & Java & Java & Java \\
Technologies of interest & Web, Mobile and Desktop & Web, Mobile and AI & Web and Mobile & Mobile and Desktop & Web, Mobile and Desktop & Web and Mobile \\
\bottomrule
\end{tabular}
\end{table*}

\subsection{Team's ICPs}
As suggested by Pinto~\cite{de2020toward}, the team's ICPs should be defined for each software development team to reflect the team's experience. Based on the collection of ICPs suggested in this work, the team's ICPs guided the team during the coding and refactoring process in the implementation phase.

\subsubsection{Defining Team ICPs}
To define the Team's ICPs, the members collaboratively established categories and weights based on their criteria, using the collection of ICPs suggested in Table ~\ref{tab:suggested_icps} as a basis. These discussions were facilitated by the first author, guiding the team in selecting or electing new ICPs. In case of disagreement, discussions were expanded to reach a consensus. Table~\ref{tab:team_icps} presents the final result.

\begin{table}[b!]
\centering
\small
\caption{Team ICPs table.}
\label{tab:team_icps}
\begin{tabular}{@{}>{\centering\arraybackslash}p{0.25\linewidth}p{0.35\linewidth}p{0.15\linewidth}p{0.10\linewidth}@{}}
\toprule
\textbf{Category} & \textbf{Description} & \textbf{subitems} & \textbf{Weight}
\\ \midrule
Branches and Loops & if, else, for, while, case and ternary & - & 1 \\
Coupling & Function as an argument and dependency on components or services & - & 1 \\
Asynchronous Function & Future and Stream & Create & 1 \\
& & Handle & 2 \\
Asynchronous widget & FutureBuilder and StreamBuilder & - & 2 \\
State Management & Using the Provider library & Notifier & 1 \\
& & Consumer & 2 \\
& Other external libraries & - & 3 \\
\midrule
\textbf{} & \multicolumn{2}{c}{\textbf{LIMIT}} & \textbf{13} \\
\bottomrule
\end{tabular}
\end{table}

When comparing the team's ICPs in Table~\ref{tab:team_icps} with the suggested ICPs collection in Table~\ref{tab:suggested_icps}, it is evident that some ICPs were not considered. For example, the ``Nullable'' and ``Animated Widget'' categories are not present in Table~\ref{tab:team_icps}. The team did not express concerns regarding the use of nullable widgets and variables. On the other hand, ``Animated Widgets'' were initially considered, but they were removed in later versions (see Table~\ref{tab:history_icps}), as the project was not as focused on UI and UX aspects.

The team designated classes as code units, with a maximum limit of 13 ICP weights per unit. Throughout the study, three versions of the ICPs were developed, being Table~\ref{tab:team_icps} the final version adopted during the Refactoring Phase.

\subsubsection{Refinement}

The refinement of the team's ICPs table was conducted through periodic analysis of the categories and weights of the ICPs by the team. This procedure was essential for the continuous evaluation and review of the items selected in the table, aligning them with the criteria established by the team members. The goal was to ensure a more accurate representation of the complexity perceived by the developers.

\subsubsection{Identification and review of ICPs}

The activities of identification and review of ICPs in the code units were carried out manually due to the lack of automatic tools for the Flutter. Team members highlighted the identified ICPs through comments in the source code, indicating the category of the ICPs and emphasizing the total weight of each code unit. 

Additionally, a pre-completion procedure was included for functionalities, with a code review conducted by the author to ensure the maintenance of CDD practices, ensuring the correct application of categories and the calculation of ICP weights determined by the team, aiming to mitigate the impacts of the lack of automated tools and the inexperience of team members.


\section{Data Collection and Analysis}

For the data collection, we focused on the Practical Project Implementation stage of the workshop. We collected data from different sources: team meetings, semi-structured interviews, and source code analysis. The details of the process is described next.

\subsection{Team meetings}

In this study, we integrated CDD activities into Scrum ceremonies. The workflow was structured to ensure continuous integration of CDD principles throughout the week. For instance:

\begin{itemize}
    \item On Mondays, refinement meetings were held to update the team's ICPs. 
    \item From Tuesday to Friday, a 15-minute daily meeting addressed CDD-related issues, ensuring that any challenges or questions could be promptly resolved. 
    \item On Saturdays, review and retrospective meetings facilitated in-depth discussions about the week's CDD activities, allowing the team to reflect on their progress and identify areas for improvement.
\end{itemize}

Throughout the practical project, a total of 26 remote meetings were held, with 91 minutes dedicated solely to CDD topics. 

To promote reflection among team members during Scrum retrospective meetings, analysis of the use of CDD and the team's ICPs table throughout the current sprint was encouraged. To achieve this, the following questions related to the ongoing sprint were formulated: 1) How do you perceive the adoption of CDD in the project?; 2) Do you perceive that the implementation of CDD is effectively contributing to making your classes more readable and less complex?; 3) Was there a need to refactor due to cognitive overload of the classes in this Sprint?; 4) What suggestions do you have to improve the team's ICPs table or adjust the ICPs weight limit that was previously established?; and 5) Do you have any comments or suggestions about the use of CDD in the project?

These questions were formulated with the purpose of generating critical insights among team members, and creating an environment conducive to identifying opportunities for improvement and optimization in the development process. Subsequently, discussions were conducted to implement the necessary changes to the team's ICPs, aligned with the team's decisions. 
This approach ensured that CDD principles were consistently applied and reviewed, enhancing the team's understanding and implementation of these practices.
Although most meetings were recorded for future reference and analysis, some were not recorded due to the presence of sensitive information or to ensure the comfort of participants. This approach balanced the need for documentation with the importance of maintaining a respectful and open environment for team members.

\subsection{Semi-structured Interviews}\label{sec:interviews}

Semi-structured interviews were conducted with team members to explore the impacts of CDD throughout the development process of the practical project, from the perspective of these members, aiming to gain a deeper understanding.

\subsubsection{Preparation}
An interview script was developed with questions subdivided by themes. A pilot interview  was also conducted for necessary adjustments. Five interviews were carried out, covering all team members who actively participated in all stages of the project. The recordings totaled 188 minutes, with an average of 38 minutes per interview, ranging from 28 to 54 minutes. All interviews were recorded and were conducted remotely via Google Meet, from November 21st to 25th, 2023.

To enrich the details of the interviews, participants were asked to download the three versions of the team's ICPs table. Additionally, they were recommended to revisit the tasks assigned to them and analyze the commits made by themselves and other team members. During the interviews, the team members responded to questions related to the following topics:

\begin{itemize}
    \item \textbf{Introduction:} Sharing information about the team's overall activities in the project, including an overview of goals, objectives, and the specific role of each team member in the context of CDD.
    \item \textbf{Preparation and Work:} Discussing the creation of the team's ICPs table, detailing the process of identifying and defining the ICPs, and explaining how the team prepared for and executed their development tasks using CDD principles.
    \item \textbf{Benefits:} Addressing potential advantages, impacts, and success evaluation criteria, such as improved code readability, maintainability, and reduced cognitive load on developers. Discuss how these benefits facilitate better learning and understanding of software design principles for novice developers.
    \item \textbf{Challenges:} Dealing with gaps in understanding, execution obstacles, and resistances faced by the team. Explore specific difficulties encountered in grasping CDD concepts, implementing them in the Flutter environment, and overcoming initial resistance to adopting new methodologies.
    \item \textbf{Lessons Learned:} Helping identify possible improvements in CDD activities. Reflect on the learning experiences, highlight key takeaways, and suggest adjustments to the CDD approach that could enhance its effectiveness for future projects and better support novice developers in mastering software design.
\end{itemize}

\subsubsection{Analysis}
With participants' consent, the interviews were recorded, later transcribed, and integrated into the artifacts of this work. This procedure allowed for a detailed analysis of the interviews, following these steps:
Initially, the interviews were watched and, when necessary, reviewed to draft a documented summary for each participant. Each team member was anonymized, identified by a previously established unique designation, such as TM1, representing team member 1. Next, relevant fragments of the interviews were identified, numbered, and categorized, documented in summaries organized by themes addressed during the interviews. Finally, a recurrence analysis of the fragments across the interviews was conducted, seeking evidence to corroborate or challenge the points discussed, contributing to a more comprehensive understanding of the results.

\subsection{Source Code Analysis}
Source code analysis was a crucial process to (1) understand the overall context via code units, (2) identify specific details through commits in the repository, and (3) relate them to the tasks of the practical project. This process allowed the collection and documentation of data for a deeper code analysis, providing a comprehensive understanding through the traceability of each context.

\subsubsection{Collection and Documentation of Metadata}
The project metadata was collected during the development and refactoring phases in the implementation of the practical project. This collection involved manual analysis of all project classes, where data such as code unit count, lines of code, and class ICPs were extracted, considering their respective weights. A total of 139 checks were performed among the project classes, with 62 at the end of the last Sprint and 77 at the end of the refactoring phase.

The obtained data was documented in a spreadsheet, organized into two tabs: ``Last Sprint,'' conducted on Oct 3rd, and ``Refactoring Phase,'' which lasted until Oct 10th. This spreadsheet provides details on the code units' compliance with the ICP limits established during these periods, offering a comprehensive view of the project's state on these specific dates.

\subsubsection{Commit Analysis}
Commit Analysis of the repository was conducted to identify evidence of events occurring during the execution of the practical project. Each task was linked to its respective commits in the repository, providing contextualization of the software development process. This approach enabled the identification of patterns, correlations, and the impacts of developers' decisions on code construction or refactoring, contributing to a deeper understanding of project progress.

Another spreadsheet was used to store these commits, organized according to the context of the features to be developed by the involved team members, along with commit metadata. In total, 34 commits were examined throughout the implementation of the practical project. This analysis was complemented by the interviews conducted (see Section~\ref{sec:interviews}) with participants, further enriching the understanding of dynamics and adjustments implemented throughout the development cycle.


\section{RQ1: How does CDD aid novice developers to understand and manage software complexity?}

To answer this question, we analyzed the ICP table, the design practices adopted during the project development, and project metadata collected. This approach allowed us to examine how CDD assisted novice developers to understanding of software design.

\vspace{0.2cm}
\noindent
\textbf{Selection and Use of ICPs.}
The selection and use of ICPs is key in helping novice developers identify the complexity within the project's source code. The development team, along with the instructors, prioritized ICPs featuring code elements that team members found challenging or unfamiliar. For example, TM1 commented: ``[The ICPs] were mostly things that I didn’t really understand how they worked.'' This illustrates that, in an educational context, the selection of ICPs can guide students toward areas they find difficult to navigate, allowing them to focus on these aspects during the learning process.

Additionally, the ICPs included code structures that could result in excessive nesting, thereby limiting the readability of the classes. One team member (TM3) mentioned: ``Anything that could bunch up a lot of code, nest a lot of code in one place. If it had the potential for that, I would include it too, even if I understand the concept.'' This practice helps newcomers identify and avoid coding practices that could lead to complex and difficult-to-maintain structures. By emphasizing these ICPs, the team was able to enhance their understanding of good coding practices and improve the overall maintainability of the codebase.

We also defined subitems for few ICPs. For instance, we slippted splitting the category ``Asynchronous Functions'' into ``Create'' and 'Handle', reflecting the perspective that handling an asynchronous function within the user interface requires more effort than creating it. This allows students to understand the varying difficulties associated with different aspects of the same functionality, facilitating deeper learning. Despite the suggested collection of ICPs by this work, it is essential for novice developers to have a clear understanding of which ICPs will be counted by the team, as indicated by CDD~\cite{de2020toward}. This encourages active participation of students in the learning and development process, enhancing their understanding and engagement.

\begin{table*}
\centering
\small
\caption{Evolution of the set of ICPs.}
\label{tab:history_icps}
\begin{tabular}{@{}>{\centering\arraybackslash}p{0.30\linewidth}p{0.20\linewidth}p{0.20\linewidth}p{0.20\linewidth}@{}}
\toprule
\textbf{ICP Category} & \textbf{Version 1} & \textbf{Version 2} & \textbf{Version 3}
\\ \midrule
Branches and Loops & Added with weight 1 & - & - \\
Basic Widget & Added with weight 1 & Removed & - \\
Coupling & Added with weight 2  & - & Changed to weight 1 \\
Asynchronous Function & Added with weight 3 & - & - \\
Asynchronous widget & Added with weight 2 & - & - \\
Nullable & Not considered & - & - \\
State Management & Added with weight 3  & - & - \\
Animated Widget & Added with weight 1  & - & Removed \\
\midrule
\textbf{LIMIT} & \textbf{32} & \textbf{30} & \textbf{13} \\
\bottomrule
\end{tabular}
\end{table*}

Finally, Table~\ref{tab:history_icps} presents the change history of the ICPs,  across three versions. 
This table illustrates the adjustments made to the ICP categories and their respective weights to better align with the project's evolving complexity management needs.
In Version 1, several ICP categories were introduced with specific weights: "Branches and Loops" with a weight of 1, "Basic Widget" with a weight of 1, "Coupling" with a weight of 2, "Asynchronous Function" with a weight of 3, "Asynchronous Widget" with a weight of 2, "State Management" with a weight of 3, and "Animated Widget" with a weight of 1. "Nullable" was not considered in this version. In Version 2, the "Basic Widget" category was removed, while the other categories remained unchanged. By Version 3, "Coupling" was adjusted to a weight of 1, and "Animated Widget" was removed, while the other categories and their weights stayed the same. The overall limit of ICPs also evolved, starting at 32 in Version 1, decreasing to 30 in Version 2, and significantly dropping to 13 in Version 3.

\vspace{0.2cm}
\noindent
\textbf{Understanding and mitigating complexity.}
As part of the project implementation, the novice developers undertook a refactoring phase to align the existing project with CDD guidelines and reduce code complexity. During this phase, the team established a stricter limit on the number of ICPs allowed in project classes, as they found that the previous limits were no longer significantly impacting code complexity. This practice also helped assess the team's learning curve; the decreasing weight of ICPs demonstrated their progress in development skills. This suggests that instructors could regularly review and adjust code quality criteria to match students' evolving profiles.

During the Refactoring Phase, team members adopted strategies such as class composition through componentization, focusing on ICPs they had practiced extensively, which facilitated the refactoring process. TM1 explained: ``I would componentize any widget that was too complex. I would divide it into parts. When it was a matter of basic programming logic, like a for loop... I would try to put it inside a separate function.'' This practice helps students understand the importance of breaking down complex problems into smaller, more manageable parts, a fundamental concept in software design.

For instance, we highlight the commit made by TM1 during this phase. The \texttt{ProjectDetails} class initially accumulated a total of 27 ICPs, exceeding the established limit. To reduce this complexity, the developer modularized the class, bringing the number of ICPs down to 13 and creating four other classes in the process. This example illustrates how strategic refactoring can significantly reduce code complexity and improve overall code quality.

\begin{figure}
    \centering
    \includegraphics[width=\linewidth]{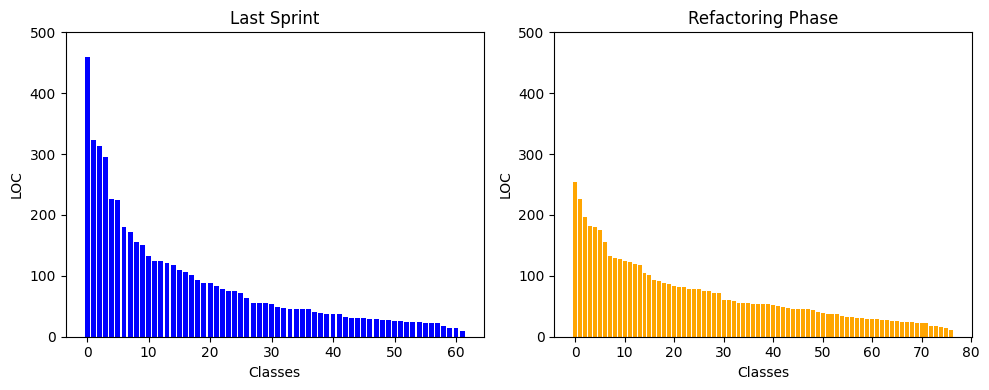}
    \caption{Distribution of LOC among project classes.}
    \label{fig:loc}
\end{figure}

\vspace{0.2cm}
\noindent
\textbf{Mitigating software complexity.}
The results of the CDD-guided refactoring, as evidenced in Figure~\ref{fig:loc}, showed a 20\% reduction in the overall average number of lines of code across project classes. Most of the refactorings involved componentization to foster reuse, which also led to an increase in the number of classes. Specifically, after the CDD-guided refactoring, there was a 24\% increase in the total number of classes.
Regarding the size of the classes, during the development phase, the average number of lines of code among the top three largest classes was 347, while the overall project average was 86 lines. After the CDD-guided refactoring, the average size of the top three classes decreased to 226 lines, and the project's average reduced to 69 lines, indicating reductions of 35\% and 20\%, respectively.
These data highlight the effectiveness of CDD-guided refactoring in reducing the size and complexity of code units. Additionally, it helped novice developers identify and address points of complexity that warranted refactoring.


\section{RQ2: How do novice developers perceive the effectiveness of CDD?}
The analysis of semi-structured interviews and the code submitted to the repository revealed several ways in which CDD contributes to improving code quality, readability, and maintainability, in particular for novice developers.

\vspace{0.2cm}
\noindent
\textbf{Project Architecture Assistance:} The CDD was fundamental in organizing the project, especially when the refactoring phase was carried out. Even without a formal architecture, CDD allowed novice developers to structure the project into more manageable modules and components. This process can help in understanding good software design practices. As TM1 mentioned, ``Even though we weren’t using any type of architecture for the project, CDD, by limiting the complexity of the classes, ended up making it a bit more organized and structured.'' By imposing limits on cognitive complexity, CDD helped novice developers to have a better understanding of the importance of well-planned and modular architecture, facilitating the assimilation of essential software design concepts.

\vspace{0.2cm}
\noindent
\textbf{Code Attention Agent:} The CDD acted as a code attention agent, guiding novice developers to focus on software design aspects. Even without a detailed understanding of other approaches, novices developers were encouraged to consider code readability and maintainability by other team members. As TM5 pointed out, ``CDD helps you stop and think if when you deliver the project, when it’s finished, any other code reviewer or the person responsible for future maintenance will be able to read it in an easier, more effective way... It's like an agent that keeps bothering you to want to improve." This constant focus on code readability and maintenance is an educational aspect that can help novices internalize coding best practices right from the start of their training.

\vspace{0.2cm}
\noindent
\textbf{Improving Code Quality and Readability:} The implementation of CDD resulted in the creation of code of better quality and more readable, as perceived by the students. As highlighted by TM5, "We took care to ensure that people would understand our code [...] It’s not enough for just the person programming the code to understand, and another person to look at the code and not understand it." These points are beneficial for the software's sustainability and especially advantageous for novice developers, who are able to learn and practice creating structured and easily understandable code. The focus on code readability and quality contributes to establishing a foundation for future software development practices.

\vspace{0.2cm}
\noindent
\textbf{Constant Refinement of Team ICPs:} Continuous optimization of the team's ICPs collection was essential to address the complexity of the code. For novice developers, this process of regular assessment and adjustment served as an educational practice, aiding in the understanding and application of Clean Code principles, and facilitating the evaluation of their learning curve. Weekly meetings were essential for adjusting coding practices and reviewing ICPs as Flutter understanding evolved. As highlighted by TM1, "I think having this meeting every week to check the table was very useful, since our level of understanding of Flutter was changing every week." This continuous process of review and learning helped newcomers adapt quickly to best development practices, fostering a dynamic and interactive learning environment.

\vspace{0.2cm}
\noindent
\textbf{Positive Acceptance by the Team:} The CDD approach was well received by the team of novice developers, who demonstrated a positive and proactive attitude. Their acceptance of CDD and adaptation to additional tasks, even in the absence of some automation tools, showed the newcomers' willingness to learn and implement new methodologies. This positive acceptance is an important indicator of CDD's potential as an effective teaching methodology, facilitating the integration of new concepts and practices into the developer training curriculum.\newpage

\section{RQ3: Challenges for new developers in understanding CDD effectively?}

In our final RQ, we list some of the challenges that novice developers faced when using CDD in an academic setting.

\vspace{0.2cm}
\noindent
\textbf{Low Impact in Early Stages:} At the beginning of the project, the use of CDD had little impact due to difficulties in integrating CDD activities along with code implementation tasks. Understanding ICPs like ``coupling" was confusing for new developers. TM1 highlighted that, "At the beginning, I had more difficulty understanding what should and should not be considered. The complexity points of each thing." Additionally, the influence of CDD was underestimated due to the lack of complexity in the classes. Moreover, the absence of clear references for the language resulted in less restrictive limits, leading to few refactorings in the early sprints. As TM2 mentioned, "It was more towards the end that we were able to apply CDD more because during the project, we hardly ever exceeded it. So it was more towards the end that the limit dropped a lot, and we had to cut out a lot of things."

\vspace{0.2cm}
\noindent
\textbf{Complementary Approaches to Effective Refactoring:} CDD is an efficient tool for novice programmers, as it requires little prior knowledge of code quality techniques. However, in learning contexts where novices do not yet have established design practices, refactoring complex units can become a challenge since they lack a robust technical repertoire to handle this complexity. An increase in the number of classes and the total ICPs in the project is observed after the refactoring period, due to the students' strategy of modularizing complex classes through class composition, dividing them into smaller units. As TM3 highlighted, "Alone, it is not enough to solve the problem." Therefore, it is important to complement the use of CDD in teaching software design by gradually introducing additional strategies that provide a more comprehensive and effective understanding in refactoring units with high complexity.

\vspace{0.2cm}
\noindent
\textbf{Manual Activities in CDD:} All CDD activities were performed manually due to the lack of supporting tools in the Flutter environment, which significantly increased the developers' effort and may have negatively impacted the project's complexity and delivery efficiency. TM1 noted, "Because it was an extra step in the work, so sometimes people ended up forgetting or were still learning how to do it. In fact, at the beginning, sometimes I didn't count correctly, but it was because I was getting used to this methodology." Additionally, there was greater effort in project management, as the responsibility for verifying project ICPs was assigned exclusively to the code reviewer during the review phase, in an effort to minimize errors. These challenges highlight the urgent need for more robust tools and processes to support methodologies like CDD, aiming to reduce manual effort and improve accuracy and efficiency in project development.

\textbf{\section{Lessons Learned}}
The implementation of CDD in an academic context has revealed several practical lessons that can assist professors and instructors when teaching software development methodologies. Integrating CDD into educational curricula ensures that beginner developers acquire essential practical and theoretical skills for high-quality software development.

\subsection{Definition of Boundaries}

\vspace{0.2cm}
\noindent
\textbf{Initial Flexibility:} Offering initial flexibility regarding the complexity limits of classes can facilitate the application of CDD with novice developers. However, it may compromise the effectiveness of practices and create gaps in software design culture. Excessive flexibility can lead to less rigorous practices, negatively impacting code quality as the software evolves. Ensuring that this flexibility is gradually reduced as students gain confidence helps them transition smoothly from learning to applying stringent practices.

\vspace{0.2cm}
\noindent
\textbf{Importance of Restriction:} Establishing strict boundaries from the start encourages the adoption of good practices and helps create less complex code, better aligning with CDD objectives and promoting a more structured development culture. This strict approach helps students internalize the importance of maintaining high standards in software design from the beginning of their education.

\subsection{Enhancement and Management of ICPs}

\vspace{0.2cm}
\noindent
\textbf{Customization of ICP Collection:} Adapting the collection of ICPs to the needs and knowledge level of students is crucial for a more targeted and efficient teaching approach. Using specific ICPs for the team or project can help balance time and resources, allowing students to focus on the most relevant aspects. Tailoring ICPs ensures that students engage with concepts that are directly applicable to their current skill level and learning objectives.
    
\vspace{0.2cm}
\noindent
\textbf{Distribution of Responsibilities:} Sharing the responsibility for reviewing ICPs among all team members, rather than centralizing it, promotes a deeper understanding of design practices and encourages collaboration. Incorporating ICP review into the code review process and distributing it among the team helps improve student autonomy and understanding of best practices. This approach fosters a collaborative learning environment where students learn from each other’s insights and experiences.

\vspace{0.2cm}
\noindent
\textbf{Gradual Approach for Introducing ICPs:} Gradually introducing new ICPs to the team’s set facilitates the assimilation of complex concepts. This method could allow students to absorb and integrate concepts progressively, promoting more effective learning aligned with the group’s experience level. A phased introduction helps prevent overwhelming students and allows them to build a strong foundational understanding before tackling more advanced topics.

\subsection{Enhancing Student Learning and Curriculum Development with CDD}

\vspace{0.2cm}
\noindent
\textbf{Improvement in Self-Assessment of Learning Curve with CDD:} CDD allows students to monitor their progress and identify areas for improvement by providing clear metrics on code complexity. This facilitates reflection on practices and enhances understanding of design and maintenance concepts. Integrating these self-assessment tools into the curriculum helps students take ownership of their learning journey and continuously strive for improvement.

\vspace{0.2cm}
\noindent
\textbf{Integration of CDD into the software engineering curriculum:} Incorporating CDD into the software engineering curriculum offers a structured approach to teaching software design. Alo, by applying CDD in practical projects helps students face complexity gradually, connecting theory and practice while improving code quality. This structured approach aids students to systematically develop the skills needed for effective software design and maintenance.

\vspace{0.2cm}
\noindent
\textbf{Establishment of Regular Feedback Cycles Based on CDD:} Using CDD activities for regular feedback helps monitoring the student progress, improving teaching practices, and promoting an adaptive learning environment. Finally, regular feedback loops help instructors identify and address learning gaps promptly, ensuring that students stay on track and continuously improve their coding skills.

\section{THREATS To VALIDITY}

First, there was a high number of participants drop outs throught the study. Out of 24 initial participants, only 6 completed the practical project. The workshop took place from July to September, starting during vacation and extending into the beginning of the academic semester, which may have contributed to dropouts due to academic and personal workload. Many participants chose to focus solely on the basic Flutter training.

Second, this low participation may have introduced bias into the results, as the final group may not fully represent the diversity of skills and interests of the initial group. This could limit the generalizability of the findings. It is important to note that the results of this research reflect only the data from participants who completed the entire workshop, as only they were exposed to the approach in question. To increase participation, it is important to establish clear expectations from the outset, emphasizing the commitment required to complete the project. Additionally, adjusting the schedule to avoid conflicts with the start of the academic semester may be beneficial. Offering ongoing support to participants facing difficulties is also essential for promoting greater engagement.

Third, while CDD practices were beneficial, their applicability in different contexts or with participants of different skill levels may result in variations in observed outcomes. Therefore, we do not claim that our results generalize for other contexts with similar settings. Finally, despite the limitations mentioned, the study provides insights on the usefulness of CDD in educational contexts. The methodology was designed to maximize learning and practical application of concepts. Observations and participant feedback provide qualitative evidence supporting the conclusions.

\section{Conclusions}
This study investigated the effectiveness of CDD in helping novice developers to design and implement an academic project in Dart and Flutter.
We observed the dynamics among team members during the development and refactoring process, identifying the benefits and challenges they faced. There was a continuous adaptation of the team's ICPs, resulting in a 20\% reduction in the number of lines of code. However, we also noted a 24\% increase in the number of project classes.
Developers highlighted improvements in code readability and quality but also mentioned the challenge of relatively low impact in the initial stages of the project, as well as the need for complementary approaches to assist in the refactoring process. 

For \emph{future work}, we plan to experiment with larger samples to validate and expand upon the findings. We also plan to survey professors and instructors to understand their software design teaching practices; and then how could CDD be better incorporated in these practices.

\section{Acknowledgements}
We thank the reviewers for their helpful comments. This work is partially supported by CNPq (308623/2022-3).

\bibliographystyle{ACM-Reference-Format}
\bibliography{references}


\begin{thebibliography}{19}


\ifx \showCODEN    \undefined \def \showCODEN     #1{\unskip}     \fi
\ifx \showDOI      \undefined \def \showDOI       #1{#1}\fi
\ifx \showISBNx    \undefined \def \showISBNx     #1{\unskip}     \fi
\ifx \showISBNxiii \undefined \def \showISBNxiii  #1{\unskip}     \fi
\ifx \showISSN     \undefined \def \showISSN      #1{\unskip}     \fi
\ifx \showLCCN     \undefined \def \showLCCN      #1{\unskip}     \fi
\ifx \shownote     \undefined \def \shownote      #1{#1}          \fi
\ifx \showarticletitle \undefined \def \showarticletitle #1{#1}   \fi
\ifx \showURL      \undefined \def \showURL       {\relax}        \fi
\providecommand\bibfield[2]{#2}
\providecommand\bibinfo[2]{#2}
\providecommand\natexlab[1]{#1}
\providecommand\showeprint[2][]{arXiv:#2}

\bibitem[dar(2023)]%
        {dart_docs}
 \bibinfo{year}{2023}\natexlab{}.
\newblock \bibinfo{booktitle}{\emph{Dart documentation}}.
\newblock
\urldef\tempurl%
\url{https://dart.dev/guides}
\showURL{%
Retrieved Dec 20, 2023 from \tempurl}


\bibitem[flu(2023)]%
        {flutter_docs}
 \bibinfo{year}{2023}\natexlab{}.
\newblock \bibinfo{booktitle}{\emph{Flutter documentation}}.
\newblock
\urldef\tempurl%
\url{https://docs.flutter.dev/}
\showURL{%
Retrieved Dec 20, 2023 from \tempurl}


\bibitem[Barbosa et~al\mbox{.}(2020)]%
        {barbosa2020revealing}
\bibfield{author}{\bibinfo{person}{Caio Barbosa}, \bibinfo{person}{Anderson
  Uch{\^o}a}, \bibinfo{person}{Daniel Coutinho}, \bibinfo{person}{Filipe
  Falc{\~a}o}, \bibinfo{person}{Hyago Brito}, \bibinfo{person}{Guilherme
  Amaral}, \bibinfo{person}{Vinicius Soares}, \bibinfo{person}{Alessandro
  Garcia}, \bibinfo{person}{Baldoino Fonseca}, \bibinfo{person}{Marcio
  Ribeiro}, {et~al\mbox{.}}} \bibinfo{year}{2020}\natexlab{}.
\newblock \showarticletitle{Revealing the social aspects of design decay: A
  retrospective study of pull requests}. In
  \bibinfo{booktitle}{\emph{Proceedings of the XXXIV Brazilian Symposium on
  Software Engineering}}. \bibinfo{pages}{364--373}.
\newblock


\bibitem[Barbosa et~al\mbox{.}(2022)]%
        {barbosa2022extent}
\bibfield{author}{\bibinfo{person}{Leonardo~Ferreira Barbosa},
  \bibinfo{person}{Victor~Hugo Pinto}, \bibinfo{person}{Alberto Luiz
  Oliveira~Tavares de Souza}, {and} \bibinfo{person}{Gustavo Pinto}.}
  \bibinfo{year}{2022}\natexlab{}.
\newblock \showarticletitle{To what extent cognitive-driven development
  improves code readability?}. In \bibinfo{booktitle}{\emph{Proceedings of the
  16th ACM/IEEE International Symposium on Empirical Software Engineering and
  Measurement}}. \bibinfo{pages}{238--248}.
\newblock


\bibitem[de~Souza and Pinto(2020)]%
        {de2020toward}
\bibfield{author}{\bibinfo{person}{Alberto Luiz Oliveira~Tavares de Souza}
  {and} \bibinfo{person}{Victor Hugo Santiago~Costa Pinto}.}
  \bibinfo{year}{2020}\natexlab{}.
\newblock \showarticletitle{Toward a definition of cognitive-driven
  development}. In \bibinfo{booktitle}{\emph{2020 IEEE International Conference
  on Software Maintenance and Evolution (ICSME)}}. IEEE,
  \bibinfo{pages}{776--778}.
\newblock


\bibitem[Evans(2004)]%
        {evans2004domain}
\bibfield{author}{\bibinfo{person}{Eric Evans}.}
  \bibinfo{year}{2004}\natexlab{}.
\newblock \bibinfo{booktitle}{\emph{Domain-driven design: tackling complexity
  in the heart of software}}.
\newblock \bibinfo{publisher}{Addison-Wesley Professional}.
\newblock


\bibitem[Ferreira(2024)]%
        {artifacts}
\bibfield{author}{\bibinfo{person}{Ronivaldo Ferreira}.}
  \bibinfo{year}{2024}\natexlab{}.
\newblock \bibinfo{booktitle}{\emph{Artifacts}}.
\newblock
\urldef\tempurl%
\url{https://drive.google.com/drive/folders/1Mv14lzTk6CqNjCwjhBn1hlQ0jcwsR_--}
\showURL{%
\tempurl}


\bibitem[Fraser et~al\mbox{.}(2003)]%
        {fraser2003test}
\bibfield{author}{\bibinfo{person}{Steven Fraser}, \bibinfo{person}{Kent Beck},
  \bibinfo{person}{Bill Caputo}, \bibinfo{person}{Tim Mackinnon},
  \bibinfo{person}{James Newkirk}, {and} \bibinfo{person}{Charlie Poole}.}
  \bibinfo{year}{2003}\natexlab{}.
\newblock \showarticletitle{Test driven development (TDD)}. In
  \bibinfo{booktitle}{\emph{International Conference on Extreme Programming and
  Agile Processes in Software Engineering}}. Springer,
  \bibinfo{pages}{459--462}.
\newblock


\bibitem[Hutton(2009)]%
        {hutton2009clean}
\bibfield{author}{\bibinfo{person}{DM Hutton}.}
  \bibinfo{year}{2009}\natexlab{}.
\newblock \showarticletitle{Clean code: a handbook of agile software
  craftsmanship}.
\newblock \bibinfo{journal}{\emph{Kybernetes}} \bibinfo{volume}{38},
  \bibinfo{number}{6} (\bibinfo{year}{2009}), \bibinfo{pages}{1035--1035}.
\newblock


\bibitem[Martin(2000)]%
        {martin2000design}
\bibfield{author}{\bibinfo{person}{Robert~C Martin}.}
  \bibinfo{year}{2000}\natexlab{}.
\newblock \showarticletitle{Design principles and design patterns}.
\newblock \bibinfo{journal}{\emph{Object Mentor}} \bibinfo{volume}{1},
  \bibinfo{number}{34} (\bibinfo{year}{2000}), \bibinfo{pages}{597}.
\newblock


\bibitem[Miller(1956)]%
        {miller1956magical}
\bibfield{author}{\bibinfo{person}{George~A Miller}.}
  \bibinfo{year}{1956}\natexlab{}.
\newblock \showarticletitle{The magical number seven, plus or minus two: Some
  limits on our capacity for processing information.}
\newblock \bibinfo{journal}{\emph{Psychological review}} \bibinfo{volume}{63},
  \bibinfo{number}{2} (\bibinfo{year}{1956}), \bibinfo{pages}{81}.
\newblock


\bibitem[Napoli(2019)]%
        {napoli2019beginning}
\bibfield{author}{\bibinfo{person}{Marco~L Napoli}.}
  \bibinfo{year}{2019}\natexlab{}.
\newblock \bibinfo{booktitle}{\emph{Beginning flutter: a hands on guide to app
  development}}.
\newblock \bibinfo{publisher}{John Wiley \& Sons}.
\newblock


\bibitem[Parnas(1994)]%
        {parnas1994software}
\bibfield{author}{\bibinfo{person}{David~Lorge Parnas}.}
  \bibinfo{year}{1994}\natexlab{}.
\newblock \showarticletitle{Software aging}. In
  \bibinfo{booktitle}{\emph{Proceedings of 16th International Conference on
  Software Engineering}}. IEEE, \bibinfo{pages}{279--287}.
\newblock


\bibitem[Pereira et~al\mbox{.}(2021)]%
        {pereira2021cognitive}
\bibfield{author}{\bibinfo{person}{Jherson Haryson~A Pereira},
  \bibinfo{person}{Alberto Luiz Oliveira Tavares~de Souza}, {and}
  \bibinfo{person}{Victor Hugo Santiago~C Pinto}.}
  \bibinfo{year}{2021}\natexlab{}.
\newblock \showarticletitle{Cognitive Load Analyzer: A Support Tool for
  Cognitive-Driven Development}. In \bibinfo{booktitle}{\emph{Proceedings of
  the XXXV Brazilian Symposium on Software Engineering}}.
  \bibinfo{pages}{468--473}.
\newblock


\bibitem[Pinto and de~Souza(2023)]%
        {pinto2023cognitive}
\bibfield{author}{\bibinfo{person}{Gustavo Pinto} {and}
  \bibinfo{person}{Alberto de Souza}.} \bibinfo{year}{2023}\natexlab{}.
\newblock \bibinfo{title}{Cognitive-Driven Development Helps Software Teams to
  Keep Code Units Under the Limit!}
\newblock
\newblock
\showeprint[arxiv]{2210.07342}~[cs.SE]


\bibitem[Pinto et~al\mbox{.}(2021)]%
        {pinto2021cognitive}
\bibfield{author}{\bibinfo{person}{Victor Hugo Santiago~C Pinto},
  \bibinfo{person}{Alberto Luiz Oliveira~Tavares de Souza},
  \bibinfo{person}{Yuri Matheus~Barboza de Oliveira}, {and}
  \bibinfo{person}{Danilo~Monteiro Ribeiro}.} \bibinfo{year}{2021}\natexlab{}.
\newblock \showarticletitle{Cognitive-Driven Development: Preliminary Results
  on Software Refactorings.}. In \bibinfo{booktitle}{\emph{ENASE}}.
  \bibinfo{pages}{92--102}.
\newblock


\bibitem[Pinto. and {Tavares De Souza}.(2022)]%
        {iceis22}
\bibfield{author}{\bibinfo{person}{Victor Hugo Santiago~C. Pinto.} {and}
  \bibinfo{person}{Alberto Luiz~Oliveira {Tavares De Souza}.}}
  \bibinfo{year}{2022}\natexlab{}.
\newblock \showarticletitle{Effects of Cognitive-driven Development in the
  Early Stages of the Software Development Life Cycle}. In
  \bibinfo{booktitle}{\emph{Proceedings of the 24th International Conference on
  Enterprise Information Systems - Volume 2: ICEIS}}. INSTICC,
  \bibinfo{publisher}{SciTePress}, \bibinfo{pages}{40--51}.
\newblock
\showISBNx{978-989-758-569-2}
\showISSN{2184-4992}
\urldef\tempurl%
\url{https://doi.org/10.5220/0011009000003179}
\showDOI{\tempurl}


\bibitem[Sweller(1988)]%
        {sweller1988cognitive}
\bibfield{author}{\bibinfo{person}{John Sweller}.}
  \bibinfo{year}{1988}\natexlab{}.
\newblock \showarticletitle{Cognitive load during problem solving: Effects on
  learning}.
\newblock \bibinfo{journal}{\emph{Cognitive science}} \bibinfo{volume}{12},
  \bibinfo{number}{2} (\bibinfo{year}{1988}), \bibinfo{pages}{257--285}.
\newblock


\bibitem[Sweller(2010)]%
        {sweller2010element}
\bibfield{author}{\bibinfo{person}{John Sweller}.}
  \bibinfo{year}{2010}\natexlab{}.
\newblock \showarticletitle{Element interactivity and intrinsic, extraneous,
  and germane cognitive load}.
\newblock \bibinfo{journal}{\emph{Educational psychology review}}
  \bibinfo{volume}{22} (\bibinfo{year}{2010}), \bibinfo{pages}{123--138}.
\newblock


\end{thebibliography}

\end{document}